# Data augmentation for battery materials using lattice scaling

Eduardo Abenza,[1,*] César Alonso,[1,*] Isabel Sobrados,[2,3] José M. Amarilla,[2,3] Javier L. Rodríguez,[1] José A. Alonso,[2,3] Roberto G.E. Martín,[+,1] and Maria C. Asensio[+,2,3]

[1] Department of Artificial Intelligence, HI-Iberia, 28036 Madrid, SPAIN
[2] Materials Science Institute of Madrid (ICMM/CSIC), Cantoblanco, E-28049 Madrid, SPAIN.
[3] MATINÉE, the CSIC Associated Unit between the Materials Science Institute (ICMUV) and the ICMM, Cantoblanco, E-28049 Madrid, SPAIN.

**ABSTRACT**

A significant step forward in Lithium-ion batteries (LIBs) developments can only be achieved by proposing mold-breaking research based on selecting the best materials for the cell components, optimizing cell manufacture, anticipating the degradation mechanisms of the LIBs, and consolidating the regeneration processes of damaged batteries. LIBs with longer recycling life, better safety, and the ability to be reused will establish sustainable state-of-the-art batteries with maximum energy efficiency, low costs, and minimal $CO_2$ emissions within a circular economy, promoting sustainability in areas as relevant as electromobility and portable electronics. Recently, there has been increasing interest in applying Artificial Intelligence (AI) techniques and their subclasses to better predict novel materials with designed properties. This collection of methods has already obtained considerable success, having been used to predict numerous physical properties of materials. However, compared to other fields, the materials data are typically much smaller and sometimes more diverse, which undoubtedly affects the construction and effectiveness of AI models. At present, several Data Augmentation (DA) methods have been proposed in materials science based on flipping, rotating, and distorting the unit cells of materials, which have been demonstrated to be very efficient in increasing the size and quality of data. Here we present an even more effective new method of Data Augmentation based on the lattice scaling of crystal structures. In the lattice scaling DA method, the unit cell is perturbed, undergoing an increase and decrease of the unit cell volume in an isotropic or anisotropic way. This transformation is particularly pertinent for battery components since volume changes of up to 5% have been reported for the insertion-based LIBs during cycling.

*These authors contributed equally

+ Corresponding authors: María C. Asensio (mc.asensio@csic.es) and Roberto G.E. Martín (robertogemartin@hi-iberia.es)

# INTRODUCTION

Materials are the keystones of every clean energy innovation in diverse domains, such as advanced batteries, solar cells, and low-energy semiconductors, among others. As the discovery and development of new materials currently imply studies over 10 to 20 years at a very high cost, finding appropriate materials is the bottleneck of the global transition to a low-carbon future [1,2]. Recently, there has been an growing relevance in applying Artificial Intelligence (AI) and its related techniques [3–6] to better predict novel materials with designed properties. This extensive collection of AI methods inspired by the outstanding success in text and image treatment is quickly and effectively spreading in materials science and engineering. Notably, in the field of materials development, considerable achievement has been attained in the context of the Materials Genome Initiative (MGI) [7,8]. The goal to find optimized materials that are highly performant drives the development and application of well-accomplished and cost-effective Machine and Deep learning (ML and DL) [9] algorithms able to predict materials with preselected properties and remarkable performance. Both schemes, closed-loop active learning approaches [10], and generative DL models are being developed to get stable materials with target properties and accessible synthesis methods [11].

At the same time, material data repositories, which feed and enhance these Artificial Intelligence frameworks, enrich and multiply rapidly [12]. However, the databases describing materials are generally much smaller than those in the text and image fields. [13]. Particularly in batteries, they are relatively small and more diverse than in other traditional AI domains, which undoubtedly affects the construction and effectiveness of ML and DL models. The battery's components are successfully mastered by selecting materials that need a complete physicochemical characterization and a full electrochemical assessment. Obviously, the most straightforward way to fit this fundamental problem of size and homogeneity of databases is directly to collect more real data. However, that is not always feasible and usually costs time and effort. In addition, raw data in materials science, mainly experimental raw data, are typically vulnerable to noise, corruption, missing values and frequently suffer from inconsistent inputs. Hence, the issue related to missing data and the significant differences in the data type is always present, as the information is often collected through multiple sources like a diversity of theoretical calculations and a wide variety of experimental techniques [14–18].

In this context, the development of ML and DL approaches is closely dependent on methodologies able to perform compelling data pre-processing and data augmentation (DA) [19,20], as poor datasets can essentially affect the accuracy and lead to incorrect predictions. That makes crucial the utilization of procedures that ensure a robust quality of the datasets. Moreover, another source of poor performance that can affect ML and DL models is the training data overfitting, which impairs the generalization ability of the models. DA is known to be a powerful technique to avoid overfitting and improving the performance of the models on unseen data [21]. Effectively, DA procedures can generate data for ML and DL models, decreasing the dependency of these methods on the size of the training data and improving the performance and precision of their results. Recently, several DA methods have been proposed in materials science based on flipping, rotating, and distorting the unit cells of materials, which have demonstrated to be very efficient at increasing the size and quality of the data [22]. They are considered an efficient way to perform data augmentation treatments without greatly distorting the original data, allowing a better-automated knowledge capture from the dataset by the AI algorithms, representing, processing, and learning from data.

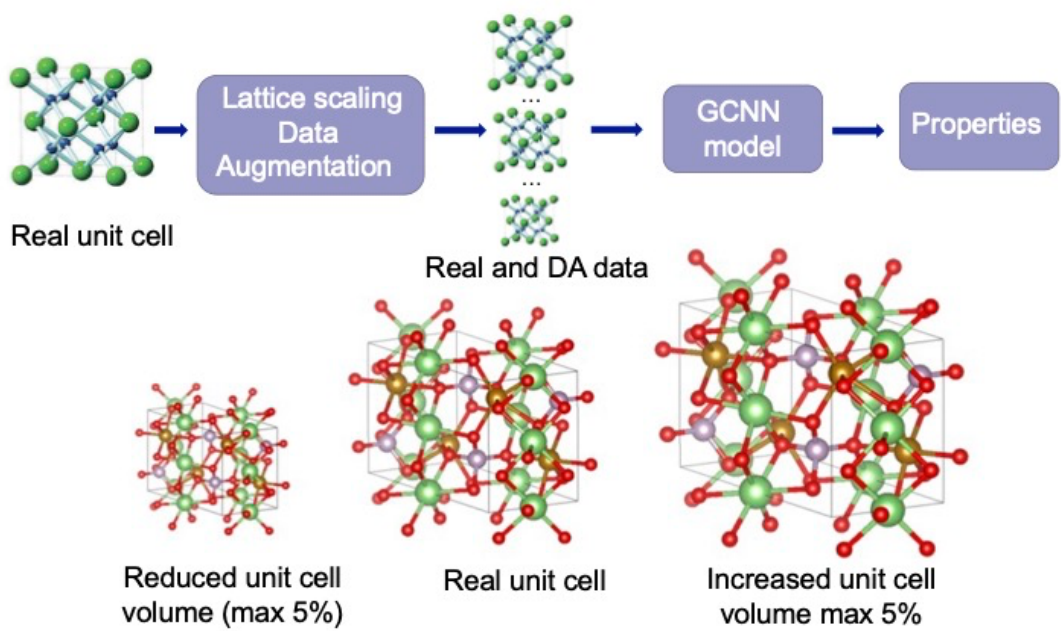

*Figure 1*: *General framework describing the lattice scaling data augmentation method and the pipeline using this data augmentation for the predictive CGCNN model.*

In this paper, we present a new method of Data Augmentation based on the lattice scaling of crystalline structures. In the lattice scaling DA method, the unit cell is perturbed by increasing and decreasing its volume in isotropic and anisotropic ways. This transformation is particularly pertinent for battery components since volume changes up to 5% have been reported for insertion-based rechargeable LIBs during the cycling processes [23] [24,25]. This DA can be used as a simple plug-in to traditional ML models such as Support Vector Machine (SVM) and more advanced DL architectures like Graph Neural Networks (GNNs). In this work, a simple machine learning pipeline is utilized to test the efficiency of the proposed lattice scaling DA method on a relatively small dataset of cathode active materials with different properties related to the material itself and to the performance of the battery that contains the material as its cathode and a metallic anode of the mobile ion. The Crystal Graph Convolutional Neural Network (CGCNN) model [26] is fed by the real and augmented data sets independently, and the results of both pipelines are compared. CGCNN learns properties of the materials based on their representation as a crystal graph. Each crystal graph is directly generated from the corresponding Crystallographic Information File (CIF) data of the considered material, which is a standard text file format for representing crystallographic information, promulgated by the International Union of Crystallography (IUCr).

Figure 1 shows that the chemical structures of the real data described by individual CIF files are considered as input of the pipeline, followed by the automatic operation of the lattice scaling to generate the DA data set. As indicated schematically in the figure, the CGCNN model is fed by the original chemical data and the data generated by the DA process as representative datasets of properties associated with batteries and materials battery components. The error metrics of these property predictions confirm that the lattice scaling DA method remarkably increases the effectiveness and performance of the CGCNN method when used for chemical and electrochemical properties prediction. Finally, the lattice scaling DA results are compared with other crystal structure DA methods already reported in the literature, including random perturbation, random rotation, random translation, and swap axes transformation, using precisely the same data sets and calculation procedures to reinforce the benefits, and promote the dissemination of our lattice scaling DA technique, we have made available an open-source library that can be widely and easily used by implementing a few lines of code. The package based on Python contains tutorials and example codes to test it and is available under request to the corresponding authors of this article.

**MODELING DESIGN, MOTIVATION, AND METHODS**

In traditional data augmentation methods, the main goal is to expand a training dataset by using transformations able to preserve the knowledge of class labels of the original dataset before the transformation. Experts in particular AI domains are usually very efficient at finding the right DA transformation to enhance the ML model performance. We propose one DA transformation, which can be automatically applied to a materials dataset by slightly modifying the unit cell of the materials which constitute the original dataset. Some transformations, like volume variations produced by small bond distance changes, distortion of angles, or anisotropic bond distance modifications, are widespread in crystal structures and can be spontaneously realized. Hence, DA transformation based on slightly modifying the size and shape of the unit cells seems to be an effective and physically meaningful way to increase the performance of the property prediction [27].

This work uses the properties reported at the Battery Explorer - Materials Project platform [28], which is a customized search tool for Li-ion electrode materials. This platform offers information about existing materials and predicts the properties of new ones related to cathodes and anodes of LIBs. In it, relevant properties like the voltage, capacity, energy density, specific density, voltage profile, oxygen evolution, and lithium positive ion diffusivity, among others, describe the materials. This Database mainly integrates results from high-throughput computing research based on quantum mechanics, solid state physics and statistical mechanics, and selected experiments. This approach, led by the Ceder group, is widely described in the following scientific reports, [29–33]. This work uses this database with and without the DA process to train the graph-based predictive model CGCNN. The selected input structures are materials with application in insertion-based batteries, such as metal-ion batteries, in the discharged state (i.e., where the cathode is "full" of the mobile ion). For the lattice scaling DA technique, the volume of the materials has been modified in different ranges of percentages. However, we consider that the most physically meaningful range of percentages has been from -5% to +5%. Performing the DA transformation with such a small volume variation is aimed at increasing the 'safety' of the transformation. In data augmentation contexts, the safety of a transformation has been defined as the likelihood that the transformed data retains the same label as the original data [21]. While label preservation is not guaranteed, a change of ± 5% in volume will be more likely to maintain the labels than a greater modification (*e.g.,* ± 30%). Furthermore, volume changes up to 5% have been experimentally observed in insertion-based LIBs during the charge/discharge steps of their life cycle, supporting the notion that such small volume changes can preserve the labels of the data [23] [24,25].

In symmetric lattice scaling, the volume of the unit cell has been increased or decreased randomly (within limits), but the relationships between the three lengths of the unit cell (a,b,c) are kept constant. In asymmetric lattice scaling, unit cell lengths (a,b,c) have also been increased and decreased randomly (and therefore also the volume) but changing the values of a, b, and c independently. Results are shown for two cases where the maximum limits of volume variation were chosen to be 5% and 30%.

To feed CGCNN, the crystalline systems are represented as graphs, where atoms constitute the graph's nodes. The bonds in these systems are represented by the edges between the nodes, where all atoms within an appropriate radius are considered bonded. Nodes and edges are associated with vector representations (attributes) that enhance the model training. In fact, node attributes encode chemical information about the chemical element in each node. In contrast, edge attributes encode the bond distance between the corresponding pair of atoms. The lattice scaling DA method slightly changes the interatomic distances. Hence the CGCNN graphs are only modified at the edge level, representing the bonds between atoms. Because the atomic bonds are based on the interatomic distance, some atoms might be considered bonded when they previously were not, or vice versa. Furthermore, the edge attributes will also be modified as the distance varies. These changes at the edge level are the reason why the lattice scaling method increases the amount of data present in the dataset. For each initial material, there will be multiple different crystal graphs that will be processed by the CGCNN model.

We have focussed on the results of the recent high-throughput search on cathode materials. The CGCNN model has been applied to predict the seven electrochemical properties and four material properties described in Table 1. The model has learned from ab initio computational results available in Battery Explorer. Briefly, the properties of potential cathodes for LIBs are obtained through first principles computations and high-throughput computational screening approaches described in previous reports[34,35].

*Table 1. List of the electrochemical and material properties predicted using the CGCNN model.*

| Property category | Property | Description |
|---|---|---|
| Electrochemical | Average voltage | Amount of electrical potential a battery holds. Units: V. |
| Electrochemical | Gravimetric capacity | Amount of electric charge an electrode material can store normalized by weight. Units: mAh/g. |
| Electrochemical | Gravimetric energy | The specific energy of a battery is a measure of how much energy a battery can store normalized by weight. Units: Wh/kg. |
| Electrochemical | Maximum voltage | The maximum voltage of a battery is the highest voltage among the different steps between total charge and total discharge. Units: V. |
| Electrochemical | Minimum voltage | The minimum voltage of a battery is the lowest voltage among the different steps between total charge and total discharge. Units: V. |
| Electrochemical | Volumetric capacity | Amount of electric charge an electrode material can store normalized per volume. Units: Ah/l. |
| Electrochemical | Volumetric energy | Energy density of a battery is a measure of how much energy a battery can store normalized per volume. Units: Wh/l. |
| Material | Formation energy | Energy of the material with respect to standard states (elements), normalized per atom. Units: eV/atom. |
| Material | Energy | Potential energy of the material, normalized per atom. Units: eV/atom. |
| Material | Fermi energy | Energy required to add an electron to the material. Units: eV. |
| Material | Band gap energy | Energy difference between the top of the valence band and the bottom of the conduction band in the electronic structure of the material. Units: eV. |

## DATASETS

The initial dataset was formed by 4401 electrode active materials for metal-ion batteries, obtained from the Battery Explorer of the Materials Project database on 2021-12-21. To train a CGCNN model, this initial dataset is split into a training set, validation set and test set. The training set, composed by 80% of the materials of the initial set, is used to directly train the model. The validation set, composed by 10% of the materials, guides the training of the model. Finally, the remaining 10% of the materials form the evaluation set, which will be used for the final evaluation of the trained model's performance.

For each data augmentation technique, the initial training set was augmented, generating a new augmented training set. In each dataset, the initial training set was repeated 10 times, and the materials were subjected to the pertinent DA transformation. This results in a dataset in which, apart from the original materials present in the initial training set, there are ten new materials for each original material. The validation and evaluation sets remained non-augmented, encouraging a fairer evaluation of the technique.

Four datasets have been generated for the lattice scaling augmentation technique.
- First, two datasets were generated with randomly sampled volumes between 95 and 105% of the volume of the original material. Regarding these two datasets, transformations were applied in both an isotropic and an anisotropic fashion. In isotropic augmentations, all lattice lengths change in the same proportion, while in anisotropic transformations the lattice lengths vary in different proportions.
- The remaining two datasets were generated with randomly sampled volumes between 70 and 130% of the volume of the original material. Transformations were also applied in an isotropic and anisotropic manner.

All lattice scaling transformations were applied using the Python library "pymatgen"[36].

In summary, the size of the training, validation, and evaluation sets before and after the lattice scaling DA is indicated in Table 2.

*Table 2. Overview of the datasets used as input for property prediction using the CGCNN model, with and without Data Augmentation. Augmented materials are used only for training.*

| Datasets | Training dataset Nº of materials | Validation dataset Nº of materials | Evaluation dataset Nº of materials |
|---|---|---|---|
| **Initial Dataset** | 3515 | 439 | 440 |
| **After Data Augmentation** | 38665 | 439 | 440 |

To compare the efficiency of the lattice scaling DA method presented in this work, we have carried out similar experiments from the same initial dataset described in Table 2, but employing four additional state-of-the-art DA transformations for crystalline chemical structures. These four DA transformations were (1) Random Perturbation, (2) Random Rotation, (3) Random Translate, and (4) Swap Axes. The same splits described in Table 2 have been used for all tested DA transformations.

These transformations, performed using the Python library "AugLiChem", are briefly described as follows:

- Random Perturbation: This augmentation transformation perturbs all the sites of the crystalline compound unit cell by a short distance between 0 and 0.5 Å. This approach is very effective because, with such a small perturbation, the main symmetry elements of the system are usually radically changed.
- Random Rotation: In the Random Rotation DA transformation, all the sites in the crystal are randomly rotated between 0 and 360 degrees.
- Random Translate: In this augmentation transformation, only a few crystal sites are displaced by a distance ranging from 0 to 0.5 Å.

Swap Axes: In this DA transformation, the coordinates of the sites in the crystal are swapped between two axes, for example, between the x- and y-axes. With this transformation, the locations of all the crystal sites are considerably displaced.

Additionally, we have also prepared a baseline dataset using oversampling, in which each instance of the initial training dataset has been repeated ten times, but without any DA transformation. The motivation of this baseline dataset is to have a dataset that is not transformed, but that contains the same amount of training samples as the augmented datasets. Therefore, the DA techniques will be compared with this baseline dataset, so that the difference in performance cannot be attributed to differences in the amount of training data employed by the CGCNN model.

In order to obtain a more robust performance evaluation, the entire process has been replicated three times for each method (e.g., for anisotropic lattice scaling 5%, three different datasets have been created). In each replicate, we have followed the procedure described above, only changing the random seed. This affects which materials belong to the training, validation, and evaluation sets; the augmentation transformations; and the training of the CGCNN model.

The complete dataset generation is described in Figure 2.

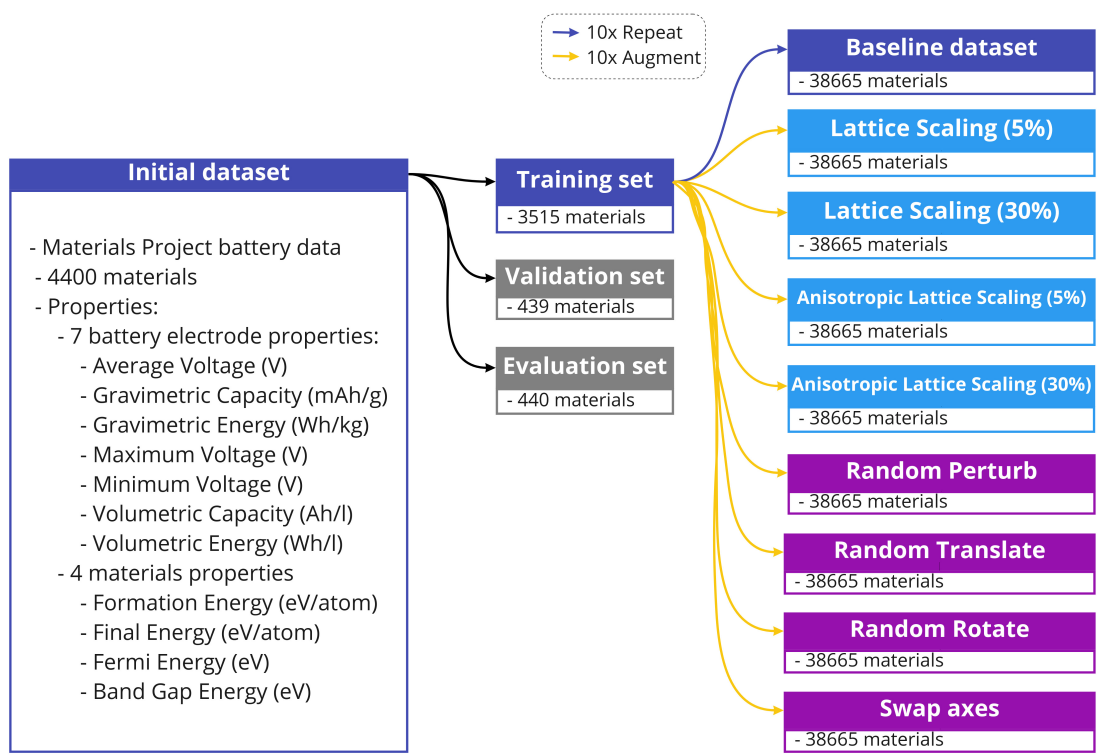

*Figure 2*: Generation of the datasets for the Data Augmentation experiments. The initial dataset is split into training, validation and evaluation sets. The training set is repeated 10 additional times to form the baseline dataset and augmented 10 times with the application of different DA transformations to create the rest of the datasets. This procedure is replicated three times in total.

**RESULTS AND DISCUSSIONS**

The enhancement in property prediction performance of the CGCNN model due to the lattice scaling DA has been evaluated by carrying out experiments to predict seven electrode materials' electrochemical properties and four material properties based on the chemical structure datasets listed in Table 2. These eleven properties have been documented in Table 1. The same procedure has been performed using the baseline dataset, and the augmented datasets. The first metric considered to evaluate the models is the Mean Absolute Error (MAE), defined as the mean of the errors (in absolute value) that the model commits when predicting the corresponding property. The lower the MAE value, the better the model behaves. In addition, the Root Mean Squared Error (RMSE) has been calculated as another performance metric. It is the square root of the mean of the errors (in quadratic value) that the model commits when predicting the respective property. A lower RMSE value indicates a better performance of the model.

Furthermore, since each property has associated different units and variance, the Mean Absolute Deviation (MAD) of each predicted property has been calculated in the evaluation set to facilitate an unbiased comparison of the behavior of the model applied to properties with different natures. MAD measures how spread out the data is in the evaluation set, similar to the standard deviation; it is the mean of the absolute difference between all the data and the mean value of the corresponding property [37]. The value of MAD represents the MAE that would be obtained by a random model that predicts the mean value of the dataset for any instance. Therefore, the MAD/MAE ratio has been evaluated, as it allows a better comparison between different properties. A model with a high MAD/MAE ratio (5 or more) is considered an excellent predictive model. Finally, all metrics and predictions have been obtained using the same CGCNN model architecture, but applying the different DA methods to the dataset.

Regarding the prediction of the electrochemical properties, Table 3 compares the lattice scaling DA methods and the previous state-of-the-art DA techniques with the baseline dataset. For the lattice scaling DA, the anisotropic lattice scaling (5%) improves the prediction of every property except for gravimetric capacity, for which this technique is neutral. The anisotropic lattice scaling (30%) improves every property prediction, except for minimum voltage, where it is neutral, and gravimetric capacity, where it worsens the performance. On the other hand, isotropic lattice scaling (5%) improves the prediction of every electrochemical property except for minimum voltage, where it is neutral, or gravimetric energy, where it performs worse than the baseline dataset. Lastly, isotropic lattice scaling (30%) only improves the predictions of gravimetric capacity, volumetric capacity and maximum voltage, worsening the predictions of the rest of the properties.

For every electrochemical property except gravimetric energy, one or multiple lattice scaling DA methods give the best MAD/MAE improvement out of all the DA techniques, performing better than the state-of-the-art augmentations. We see remarkable improvements, up to 10.5% in the case of anisotropic lattice scaling (30%) applied to maximum voltage prediction. Regarding gravimetric energy, although anisotropic lattice scaling performs better than the baseline, the best results are obtained by the random perturbation DA method.

An additional model was also trained with the initial dataset, but it consistently showed a worse performance than the other methods, due to the fact that it contained 10 times less data than the baseline or the augmented datasets.

It should be noted that, except for a few cases, the state-of-the-art DA techniques for crystalline materials perform worse than the baseline dataset, where we are simply oversampling 10 times the initial dataset without any additional transformation. This topic will be discussed later.

*Table 3. Comparison of electrochemical property prediction with different techniques of Data Augmentation, expressed as the mean MAD/MAE improvement with respect to the baseline dataset. For MAD/MAE ratio, a higher metric is indicative of a better performance. The higher, the better. For a recapitulative of the comparative metrics on the prediction of the CGCNN model for all methods (MAE, RMSE, and MAD/MAE ratio), see Tables S1-S7.*

|  | Average Voltage (V) | Gravimetric Capacity (mAh/g) | Gravimetric Energy (Wh/kg) | Maximum voltage (V) | Minimum voltage (V) | Volumetric capacity (Ah/l) | Volumetric Energy (Wh/l) |
| --- | --- | --- | --- | --- | --- | --- | --- |
| Anisotropic Lattice Scaling (5%) | 4.6% | 0.5% | 4.2% | 3.1% | **3.3%** | **7.3%** | **9.2%** |
| Anisotropic Lattice Scaling (30%) | **6.0%** | -1.1% | 6.9% | **10.5%** | 0.7% | 2.4% | 7.8% |
| Isotropic Lattice Scaling (5%) | 2.5% | 1.4% | -2.2% | 4.4% | -0.3% | 3.0% | 7.8% |
| Isotropic Lattice Scaling (30%) | -3.0% | **1.8%** | 0.6% | 2.4% | -5.0% | 4.8% | -1.5% |
| Random Translate | -1.8% | -3.8% | -6.1% | -2.4% | -1.9% | 1.6% | 2.7% |
| Random Perturb | -3.4% | -7.8% | **10.7%** | 9.3% | -0.2% | -0.6% | 4.3% |
| Random Rotate | -8.1% | -2.5% | -3.1% | -0.8% | -8.9% | 0.7% | 1.8% |
| Swap Axes | -7.5% | -1.0% | -7.2% | -7.1% | -12.8% | -1.6% | -2.2% |

The same strategy has been followed to evaluate the effectiveness of the lattice scaling DA transformation on the enhancement of the materials' properties prediction using the CGCNN model.

Apart from the electrochemical properties of the electrode materials, the CGCNN models were also trained to predict four materials properties: formation energy, energy, Fermi energy and band gap energy. For formation energy, every DA technique performed worse than the baseline. For energy per atom, the impact of every DA technique was either neutral or positive, with isotropic lattice scaling (5%) providing the greatest improvement, followed by isotropic lattice scaling (30%) In the case of Fermi energy, both types of lattice scaling (5%) had a positive impact, while most of the state-of-the-art augmentation methods impaired the performance of the model. Finally, band gap energy improved the most with anisotropic lattice scaling (5%), having mixed results with the other techniques.

*Table 4: Comparison of material property prediction with different techniques of Data Augmentation, expressed as the mean MAD/MAE improvement with respect to the baseline dataset. For MAD/MAE ratio, a higher metric is indicative of a better performance. For a recapitulative of the comparative metrics on the prediction of the CGCNN model for all methods (MAE, RMSE, and MAD/MAE ratio), see Tables S8-S11.*

|  | Formation energy (eV/atom) | Energy (eV/atom) | Fermi energy (eV) | Band gap energy (eV) |
| --- | --- | --- | --- | --- |
| **Anisotropic Lattice Scaling (5%)** | **-0.6%** | 0.2% | **2.6%** | **3.7%** |
| **Anisotropic Lattice Scaling (30%)** | -15.6% | -1.9% | -0.1% | 1.6% |
| **Isotropic Lattice Scaling (5%)** | -8.3% | **3.8%** | 2.2% | 0.1% |
| **Isotropic Lattice Scaling (30%)** | -1.1% | 1.8% | -3.6% | -4.0% |
| **Random Translate** | -9.2% | 0.6% | 1.6% | 1.4% |
| **Random Perturb** | -16.9% | -4.5% | -6.2% | 0.2% |
| **Random Rotate** | -8.0% | -0.8% | -3.2% | -4.5% |
| **Swap Axes** | -4.1% | -2.9% | -4.8% | -5.4% |

In summary, lattice scaling DA, and especially anisotropic lattice scaling with a maximum volume change of 5%, generally improve the performance of the CGCNN predictive model in comparison to the baseline dataset and to the other state-of-the-art DA techniques (with some exceptions).

In most of the cases, the state-of-the-art DA techniques tend to perform worse than the baseline dataset. In this study, we have used the default configuration for the different DA methods, which could be one of the reasons for this substandard performance. Using hyperparameter tuning to select the best configuration for each technique could be essential in improving their performance in material and electrochemical property prediction. Likewise, this tuning could be applied to our lattice scaling method, to further increase the predictive performance of the CGCNN model.

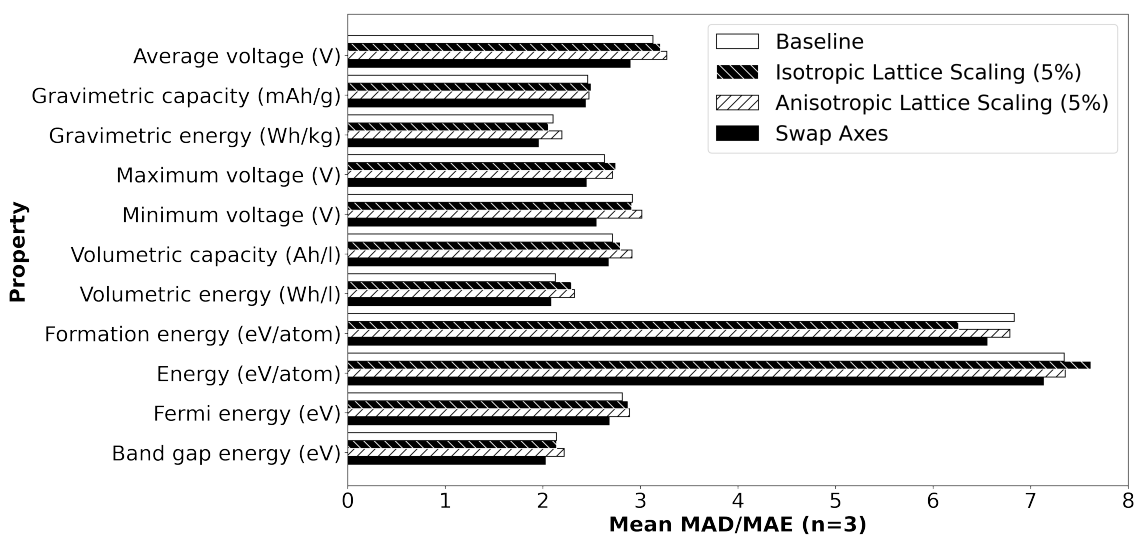

*Figure 3*: Comparison of the mean MAD/MAE ratio between isotropic and anisotropic lattice scaling (5%), the Swap Axes DA transformation, and the baseline dataset.

It should be noted that the performance of each DA technique seems to be related to the property that is being predicted. The prediction of some material properties, such as formation energy, is always impaired when using data augmentation, independently of the transformation. On the contrary, properties such as volumetric energy improve with almost every DA transformation when comparing it to the baseline. By applying a data augmentation transformation while maintaining the same labels, we are biasing the model into considering that this particular transformation does not affect the predicted property. For instance, in the case of formation energy, we hypothesize that the volume of the unit cell and, in general, atomic displacements, affect the final value of the property. Therefore, when applying any DA technique, the prediction would not improve on this property, because it depends on the parameters that we are modifying. The same reasoning could be applied to the other materials and electrochemical properties.

Furthermore, the initial dataset from Materials Project contains electrode active materials of different metal-ion batteries. A recent work on image data augmentation shows that certain augmentation techniques can enhance the prediction on certain classes, while impair the prediction on other classes [38]. In this work, we evaluate the augmentation techniques without distinguishing between the different inserted metal ions. However, it is known that insertion-based batteries with different metal ions experiment different volume changes during their charge and discharge steps, as seen in the Materials Project database [39–42]. Given the difference in volume changes, and the fact that the lattice scaling works by modifying the volume of the materials, it would be interesting to analyze the performance of this technique on the batteries with different mobile ions, to evaluate whether it depends on the type of ion. This knowledge could be useful for practical applications, to choose the method that better improves the prediction on LIBs instead of metal-ion batteries in general.

In conclusion, lattice scaling is a powerful technique to improve the prediction of some electrochemical and material properties. It could have many possible near-future applications in predicting electrochemical properties, as it can ensure substantial gains over other standard DA approaches, given that, with some exceptions, the standard state-of-the-art DA transformations produce a worse performance than simply oversampling the initial dataset.

## DATA AVAILABILITY

The data that support this work are available in the article and Supplementary Information file. Further raw data can be found at the CPE database: https://smartmaterial.hi-iberia.es/app/account/login/?next=/app/ and are available from the corresponding author (RGEM or MCA) upon requests

**ACKNOWLEDGEMENTS**

The present research has been undertaken in the context of the Associated Research Unit MATINÉE of the CSIC (Spanish Scientific Research Council), created between the Institute of Materials Science (ICMUV) of Valencia University and the Materials Science Institute of Madrid (ICMM). The authors acknowledge financial support from the MIG-20201021, LION-HD industrial research of strategic materials for lithium-ion batteries with high energy density and cost optimization in sustainable electromobility. Project granted by the Program of "MISIONES" of the CDTI, within the framework of the State Program for Business Leadership in I+D+I, of the Spanish Scientific and Technical Research Strategy plan for the Innovation 2017-2020. The Research Funds of the Valencian Community, through the project PROMETEO/2020/091 entitled "Heterostructures of two-dimensional materials and topological materials for clean and efficient energy and secure communications." Also, the project MAT2020 NIRVANA (PID2020-119628RB-C32), entitled "Metal halide perovskites: new structures for new challenges," was financed by the Ministry of Science and Innovation, Government of Spain. Finally, the present work has been supported by the project INGENIOUS (TED2021-132656B-C21 & TED2021-132656B-C22) entitled "Lithium-ion batteries: an effective methodology for the challenge of the optimization and regeneration of their main components," granted by the Call 2021 - "Ecological Transition and Digital Transition Projects." Promoted by the Ministry of Science and Innovation, Funded by the European Union within the "NextGeneration" EU program, the Recovery, Transformation, and Resilience Plan, and the State Investigation Agency. HI-IBERIA has fully supported the computing work.

**AUTHOR CONTRIBUTIONS**

EA, CA and MCA conceived the idea and designed the calculations. MCA wrote the first draft of the paper, which was enriched with further contributions from all the authors. RGEM and MCA organized the research project, and EA and CA participated in the computational design and discussion with IS, JMA, JLR, JAA, RGEM, and MCA. All the authors were deeply involved in the analyses of input data and computational results.

**COMPETING INTERESTS**

The authors declare no competing interests.

**ADDITIONAL INFORMATION**

[Data augmentation for batteries materials_SM.pdf](Data augmentation for batteries materials_SM.pdf)

Corresponding authors: the requests for materials and information should be addressed to RGEM or MCA to  robertogemartin@hi-iberia.es and mc.asensio@csic.es.

# Supplementary Materials
# Data augmentation for battery materials using lattice scaling

Eduardo Abenza,[1,*] César Alonso,[1,*] Isabel Sobrados,[2,3] José M. Amarilla,[2,3] Javier L. Rodríguez,[1] José A. Alonso,[2,3] Roberto G.E. Martín,[+,1] and Maria C. Asensio[+,2,3]

[1]Department of Artificial Intelligence, HI-Iberia, 28036 Madrid, SPAIN

[2]Materials Science Institute of Madrid (ICMM/CSIC), Cantoblanco, E-28049 Madrid, SPAIN.

[3]MATINÉE, the CSIC Associated Unit between the Materials Science Institute (ICMUV) and the ICMM, Cantoblanco, E-28049 Madrid, SPAI

*Table S1 Statistical evaluation of the Average Voltage predictions.*

|  | Mean MAD/MAE | Standard Deviation MAD/MAE | Improvement with respect to Baseline |
|---|---|---|---|
| *Anisotropic Lattice Scaling (30%)* | *3.316* | *0.070* | *6.0%* |
| *Anisotropic Lattice Scaling (5%)* | *3.272* | *0.114* | *4.6%* |
| *Isotropic Lattice Scaling (5%)* | *3.206* | *0.192* | *2.5%* |
| *Baseline* | *3.129* | *0.053* | *0.0%* |
| *Random Translate* | *3.072* | *0.135* | *-1.8%* |
| *Isotropic Lattice Scaling (30%)* | *3.035* | *0.159* | *-3.0%* |
| *Random Perturb* | *3.024* | *0.188* | *-3.4%* |
| *Swap Axes* | *2.894* | *0.141* | *-7.5%* |
| *Random Rotate* | *2.877* | *0.309* | *-8.1%* |

**Table S1. Descriptive statistics of the evaluation results obtained in the prediction of Average Voltage, grouped by dataset.**

*Table S2. Statistical evaluation of the Gravimetric Capacity predictions.*

|  | Mean MAD/MAE | Standard Deviation MAD/MAE | Improvement with respect to Baseline |
|---|---|---|---|
| *Isotropic Lattice Scaling (30%)* | *2.505* | *0.444* | *1.8%* |
| *Isotropic Lattice Scaling (5%)* | *2.496* | *0.309* | *1.4%* |
| *Anisotropic Lattice Scaling (5%)* | *2.474* | *0.382* | *0.5%* |
| *Baseline* | *2.461* | *0.371* | *0.0%* |
| *Swap Axes* | *2.437* | *0.446* | *-1.0%* |
| *Anisotropic Lattice Scaling (30%)* | *2.435* | *0.156* | *-1.1%* |
| *Random Rotate* | *2.400* | *0.287* | *-2.5%* |
| *Random Translate* | *2.367* | *0.430* | *-3.8%* |
| *Random Perturb* | *2.270* | *0.140* | *-7.8%* |

**Table S2. Descriptive statistics of the evaluation results obtained in the prediction of Gravimetric Capacity, grouped by dataset.**

*Table S3. Statistical evaluation of the Gravimetric Energy predictions.*

|  | Mean MAD/MAE | Standard Deviation MAD/MAE | Improvement with respect to Baseline |
|---|---|---|---|
| Random Perturb | 2.331 | 0.226 | 10.7% |
| Anisotropic Lattice Scaling (30%) | 2.252 | 0.201 | 6.9% |
| Anisotropic Lattice Scaling (5%) | 2.195 | 0.236 | 4.2% |
| Isotropic Lattice Scaling (30%) | 2.119 | 0.112 | 0.6% |
| Baseline | 2.106 | 0.200 | 0.0% |
| Isotropic Lattice Scaling (5%) | 2.059 | 0.238 | -2.2% |
| Random Rotate | 2.041 | 0.138 | -3.1% |
| Random Translate | 1.977 | 0.174 | -6.1% |
| Swap Axes | 1.955 | 0.090 | -7.2% |

*Table S3. Descriptive statistics of the evaluation results obtained in the prediction of Gravimetric Energy, grouped by dataset.*

*Table S4. Statistical evaluation of the Maximum Voltage predictions*

|  | Mean MAD/MAE | Standard Deviation MAD/MAE | Improvement with respect to Baseline |
|---|---|---|---|
| Anisotropic Lattice Scaling (30%) | 2.908 | 0.179 | 10.5% |
| Random Perturb | 2.877 | 0.215 | 9.3% |
| Isotropic Lattice Scaling (5%) | 2.749 | 0.166 | 4.4% |
| Anisotropic Lattice Scaling (5%) | 2.714 | 0.019 | 3.1% |
| Isotropic Lattice Scaling (30%) | 2.695 | 0.246 | 2.4% |
| Baseline | 2.632 | 0.074 | 0.0% |
| Random Rotate | 2.612 | 0.222 | -0.8% |
| Random Translate | 2.570 | 0.156 | -2.4% |
| Swap Axes | 2.445 | 0.170 | -7.1% |

*Table S4. Descriptive statistics of the evaluation results obtained in the prediction of Maximum Voltage, grouped by dataset.*

## Table S5. Statistical evaluation of the Minimum Voltage predictions

|  | Mean MAD/MAE | Standard Deviation MAD/MAE | Improvement with respect to Baseline |
|---|---|---|---|
| Anisotropic Lattice Scaling (5%) | 3.016 | 0.151 | 3.3% |
| Anisotropic Lattice Scaling (30%) | 2.940 | 0.153 | 0.7% |
| Baseline | 2.919 | 0.186 | 0.0% |
| Random Perturb | 2.914 | 0.157 | -0.2% |
| Isotropic Lattice Scaling (5%) | 2.911 | 0.058 | -0.3% |
| Random Translate | 2.864 | 0.142 | -1.9% |
| Isotropic Lattice Scaling (30%) | 2.772 | 0.266 | -5.0% |
| Random Rotate | 2.659 | 0.059 | -8.9% |
| Swap Axes | 2.546 | 0.223 | -12.8% |

Table S5. Descriptive statistics of the evaluation results obtained in the prediction of Minimum Voltage, grouped by dataset.

## Table S6. Statistical evaluation of the Volumetric Capacity predictions

|  | Mean MAD/MAE | Standard Deviation MAD/MAE | Improvement with respect to Baseline |
|---|---|---|---|
| Anisotropic Lattice Scaling (5%) | 2.914 | 0.221 | 7.3% |
| Isotropic Lattice Scaling (30%) | 2.844 | 0.226 | 4.8% |
| Isotropic Lattice Scaling (5%) | 2.796 | 0.277 | 3.0% |
| Anisotropic Lattice Scaling (30%) | 2.779 | 0.240 | 2.4% |
| Random Translate | 2.758 | 0.231 | 1.6% |
| Random Rotate | 2.734 | 0.243 | 0.7% |
| Baseline | 2.715 | 0.309 | 0.0% |
| Random Perturb | 2.698 | 0.100 | -0.6% |
| Swap Axes | 2.671 | 0.261 | -1.6% |

Table S6. Descriptive statistics of the evaluation results obtained in the prediction of Volumetric Capacity, grouped by dataset.

## Table S7. Statistical evaluation of the Volumetric Energy predictions

|  | Mean MAD/MAE | Standard Deviation MAD/MAE | Improvement with respect to Baseline |
|---|---|---|---|
| Anisotropic Lattice Scaling (5%) | 2.325 | 0.178 | 9.2% |
| Anisotropic Lattice Scaling (30%) | 2.296 | 0.239 | 7.8% |
| Isotropic Lattice Scaling (5%) | 2.294 | 0.184 | 7.8% |
| Random Perturb | 2.221 | 0.081 | 4.3% |
| Random Translate | 2.186 | 0.121 | 2.7% |
| Random Rotate | 2.167 | 0.087 | 1.8% |
| Isotropic Lattice Scaling (30%) | 2.161 | 0.137 | 1.5% |
| Baseline | 2.129 | 0.125 | 0.0% |
| Swap Axes | 2.082 | 0.181 | -2.2% |

Table S7. Descriptive statistics of the evaluation results obtained in the prediction of Volumetric Energy, grouped by dataset.

## Table S8. Statistical evaluation of the Formation Energy predictions

|  | Mean MAD/MAE | Standard Deviation MAD/MAE | Improvement with respect to Baseline |
|---|---|---|---|
| Baseline | 6.832 | 0.674 | 0.0% |
| Anisotropic Lattice Scaling (5%) | 6.788 | 0.838 | -0.6% |
| Isotropic Lattice Scaling (30%) | 6.755 | 0.747 | -1.1% |
| Swap Axes | 6.554 | 0.566 | -4.1% |
| Random Rotate | 6.283 | 0.875 | -8.0% |
| Isotropic Lattice Scaling (5%) | 6.264 | 1.301 | -8.3% |
| Random Translate | 6.205 | 1.642 | -9.2% |
| Anisotropic Lattice Scaling (30%) | 5.763 | 0.323 | -15.6% |
| Random Perturb | 5.679 | 0.497 | -16.9% |

Table S8. Descriptive statistics of the evaluation results obtained in the prediction of Formation Energy, grouped by dataset.

## Table S9. Statistical evaluation of the Energy predictions

|  | Mean MAD/MAE | Standard Deviation MAD/MAE | Improvement with respect to Baseline |
|---|---|---|---|
| Isotropic Lattice Scaling (5%) | 7.620 | 0.495 | 3.8% |
| Isotropic Lattice Scaling (30%) | 7.474 | 0.478 | 1.8% |
| Random Translate | 7.389 | 0.580 | 0.6% |
| Anisotropic Lattice Scaling (5%) | 7.356 | 0.575 | 0.2% |
| Baseline | 7.344 | 0.661 | 0.0% |
| Random Rotate | 7.287 | 0.386 | -0.8% |
| Anisotropic Lattice Scaling (30%) | 7.202 | 0.112 | -1.9% |
| Swap Axes | 7.131 | 0.225 | -2.9% |
| Random Perturb | 7.014 | 0.153 | -4.5% |

Table S9. Descriptive statistics of the evaluation results obtained in the prediction of Energy, grouped by dataset.

## Table S10. Statistical evaluation of the Fermi Energy predictions

|  | Mean MAD/MAE | Standard Deviation MAD/MAE | Improvement with respect to Baseline |
|---|---|---|---|
| Anisotropic Lattice Scaling (5%) | 2.889 | 0.029 | 2.6% |
| Isotropic Lattice Scaling (5%) | 2.876 | 0.114 | 2.2% |
| Random Translate | 2.859 | 0.114 | 1.6% |
| Baseline | 2.815 | 0.106 | 0.0% |
| Anisotropic Lattice Scaling (30%) | 2.812 | 0.114 | -0.1% |
| Random Rotate | 2.724 | 0.006 | -3.2% |
| Isotropic Lattice Scaling (30%) | 2.713 | 0.129 | -3.6% |
| Swap Axes | 2.680 | 0.188 | -4.8% |
| Random Perturb | 2.641 | 0.029 | -6.2% |

Table S10. Descriptive statistics of the evaluation results obtained in the prediction of Fermi Energy, grouped by dataset.

## Table S11. Statistical evaluation of the Band Gap Energy predictions

|  | Mean MAD/MAE | Standard Deviation MAD/MAE | Improvement with respect to Baseline |
|---|---|---|---|
| Anisotropic Lattice Scaling (5%) | 2.219 | 0.117 | 3.7% |
| Anisotropic Lattice Scaling (30%) | 2.175 | 0.147 | 1.6% |
| Random Translate | 2.171 | 0.120 | 1.4% |
| Random Perturb | 2.145 | 0.194 | 0.2% |
| Isotropic Lattice Scaling (5%) | 2.142 | 0.196 | 0.1% |
| Baseline | 2.140 | 0.126 | 0.0% |
| Isotropic Lattice Scaling (30%) | 2.054 | 0.166 | -4.0% |
| Random Rotate | 2.043 | 0.143 | -4.5% |
| Swap Axes | 2.025 | 0.088 | -5.4% |

Table S11. Descriptive statistics of the evaluation results obtained in the prediction of Band Gap Energy, grouped by dataset.

## Figure S1: Different average voltage predictions

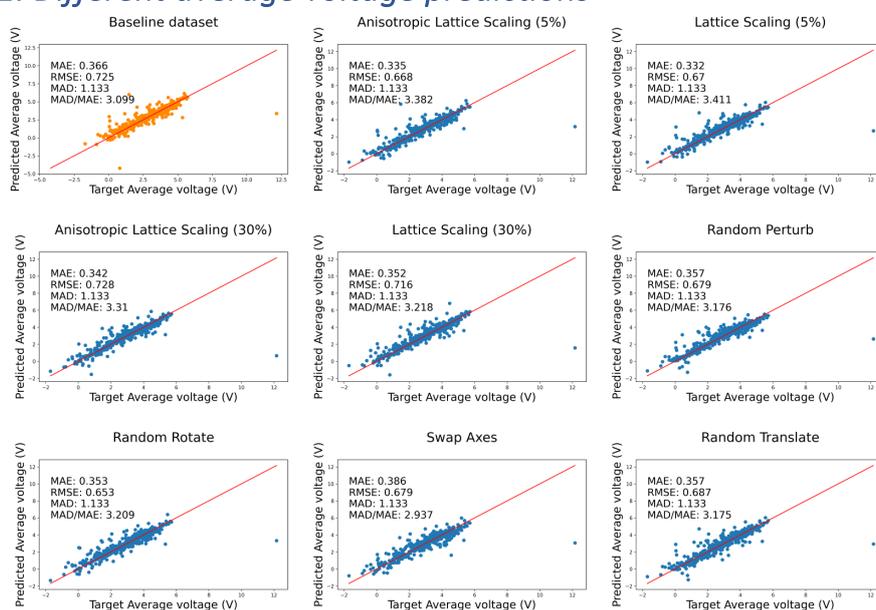

Figure S1: Comparison between different data augmentation strategies for average voltage prediction. The relationship between the predicted and real values of the property is shown.

*Figure S2: Different gravimetric capacity predictions.*

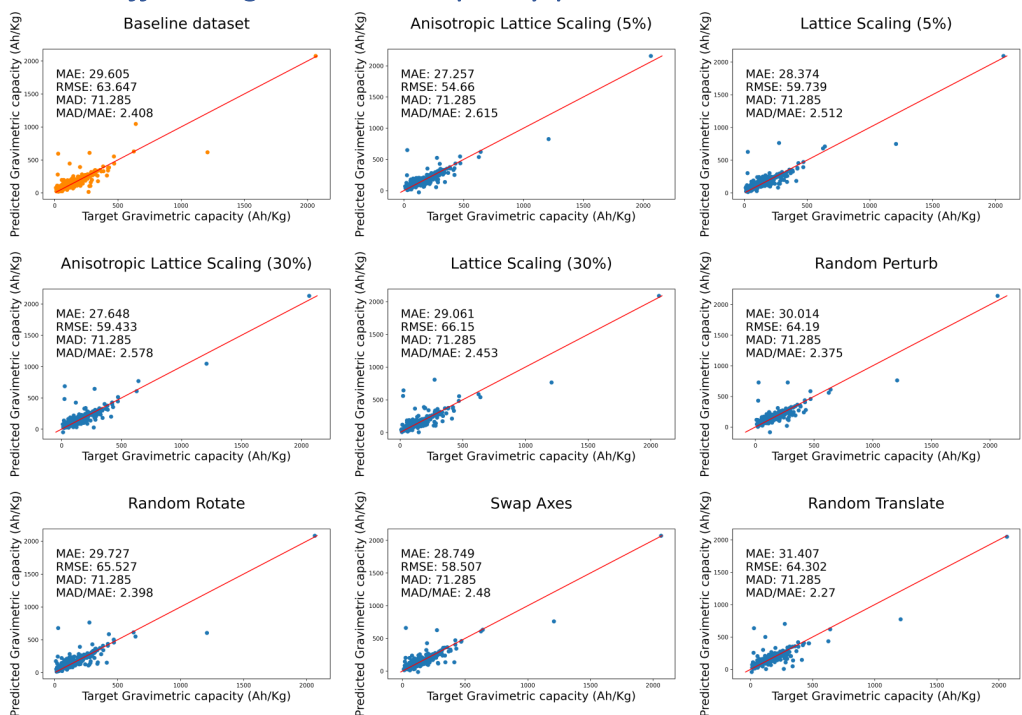

***Figure S2:** Comparison between different data augmentation strategies for gravimetric capacity prediction. The relationship between the predicted and real values of the property is shown.*

*Figure S3: Different gravimetric energy predictions.*

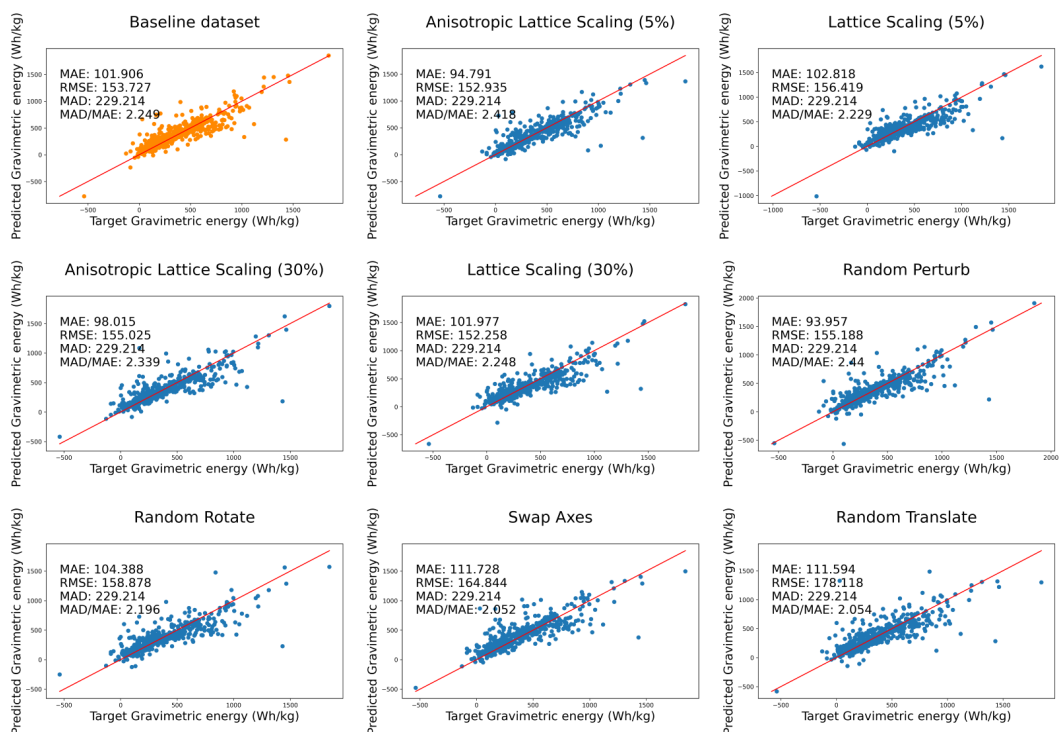

***Figure S3:** Comparison between different data augmentation strategies for gravimetric energy prediction. The relationship between the predicted and real values of the property is shown.*

*Figure S4: Different maximum voltage predictions.*

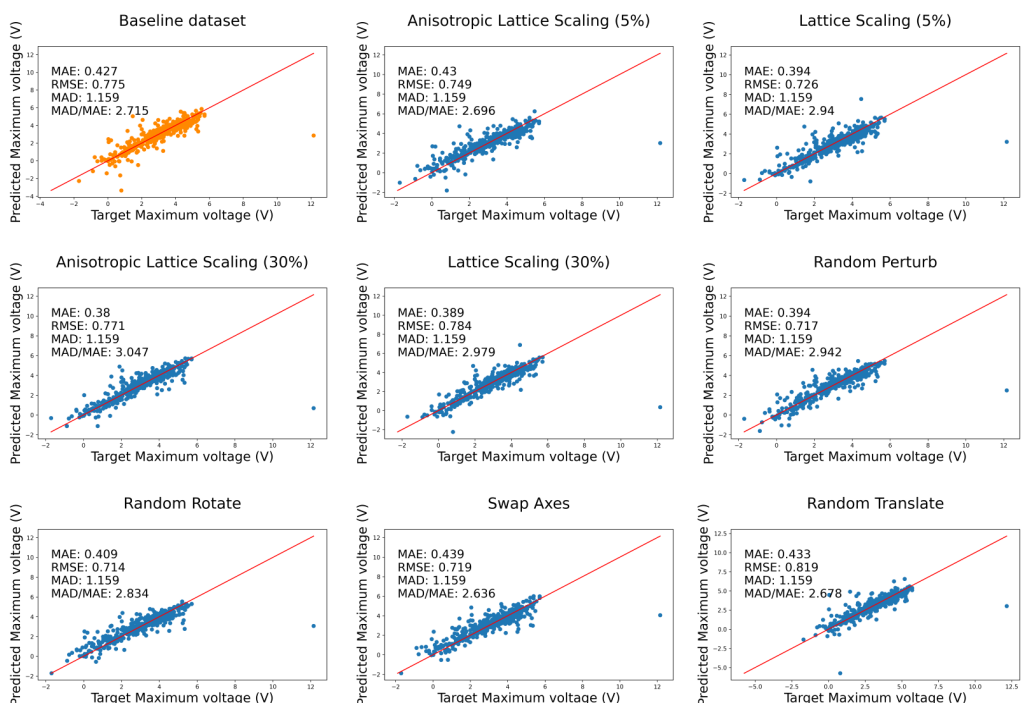

***Figure S4:** Comparison between different data augmentation strategies for maximum voltage prediction. The relationship between the predicted and real values of the property is shown.*

*Figure S5: Different minimum voltage predictions.*

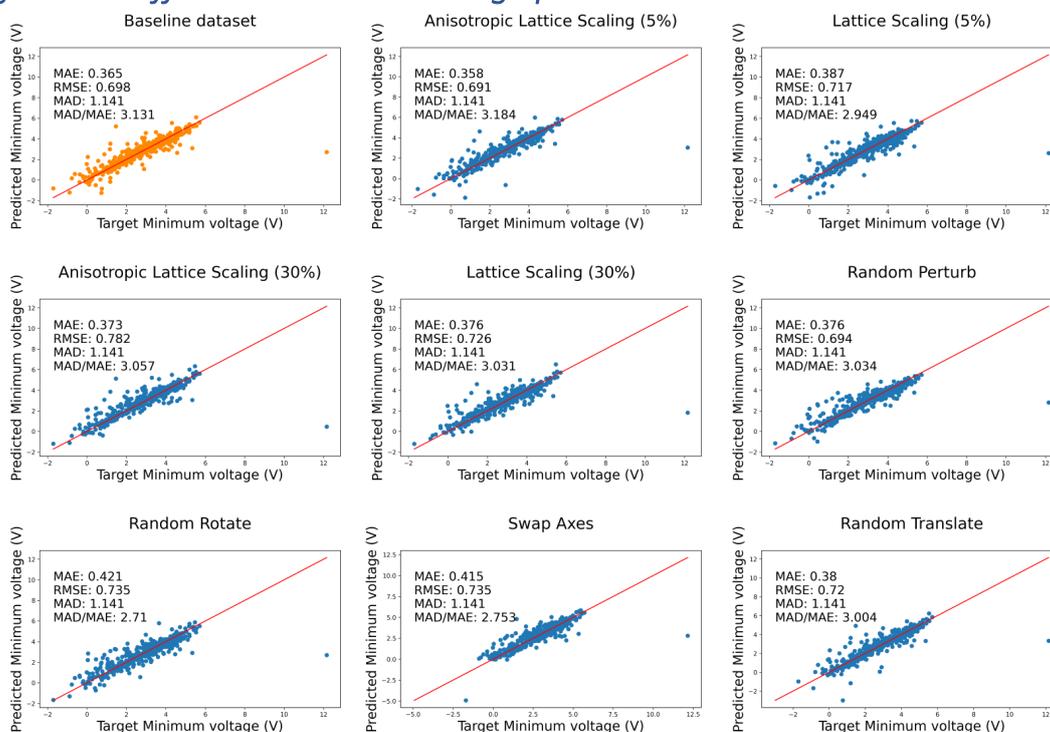

***Figure S5:** Comparison between different data augmentation strategies for minimum voltage prediction. The relationship between the predicted and real values of the property is shown.*

*Figure S6: Different volumetric capacity predictions.*

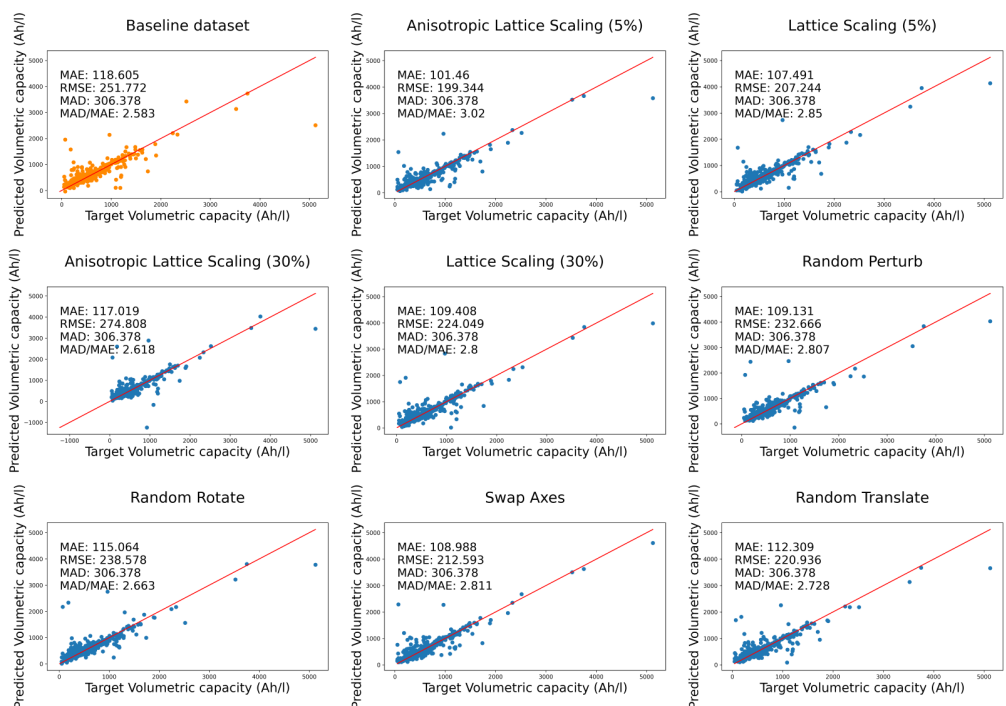

*Figure S6: Comparison between different data augmentation strategies for volumetric capacity prediction. The relationship between the predicted and real values of the property is shown.*

*Figure S7: Different volumetric energy predictions.*

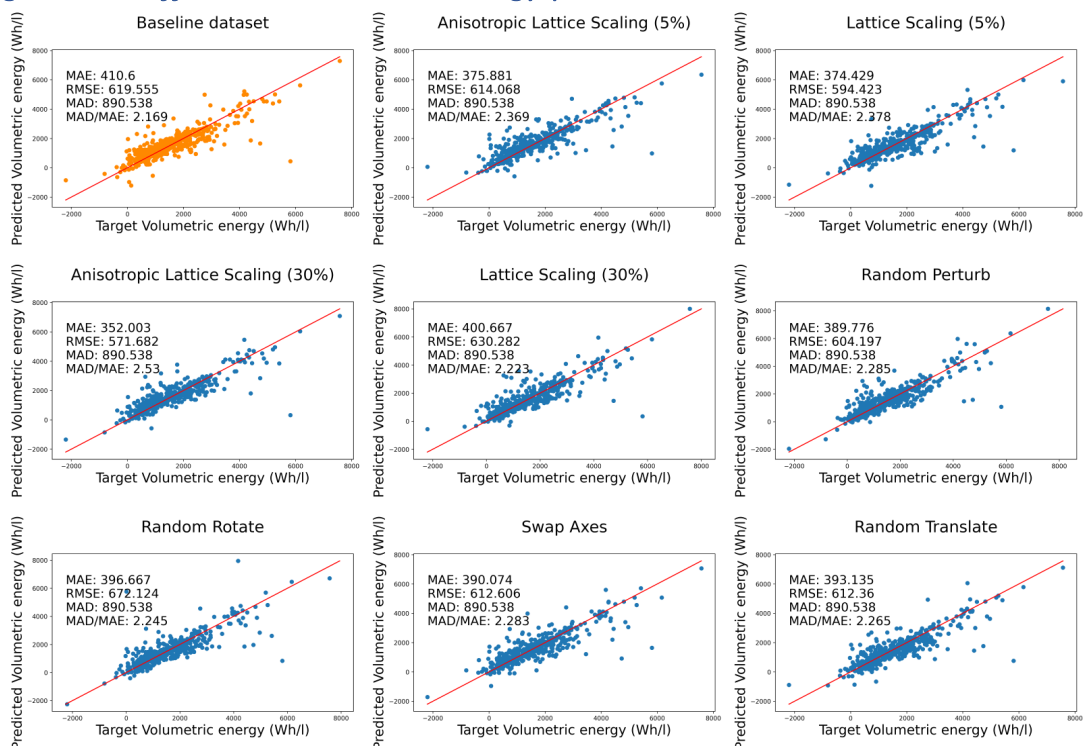

*Figure S7: Comparison between different data augmentation strategies for volumetric energy prediction. The relationship between the predicted and real values of the property is shown.*

## *Figure S8: Different formation energy per atom predictions.*

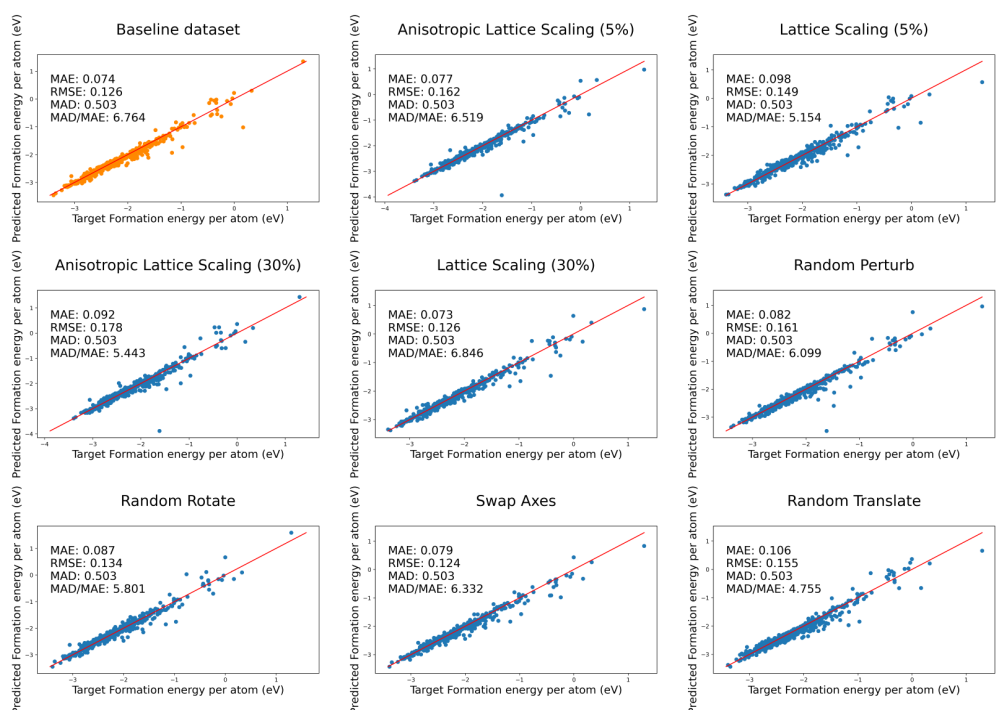

***Figure S8:** Comparison between different data augmentation strategies for formation energy per atom prediction. The relationship between the predicted and real values of the property is shown.*

## *Figure S9: Different energy per atom predictions*

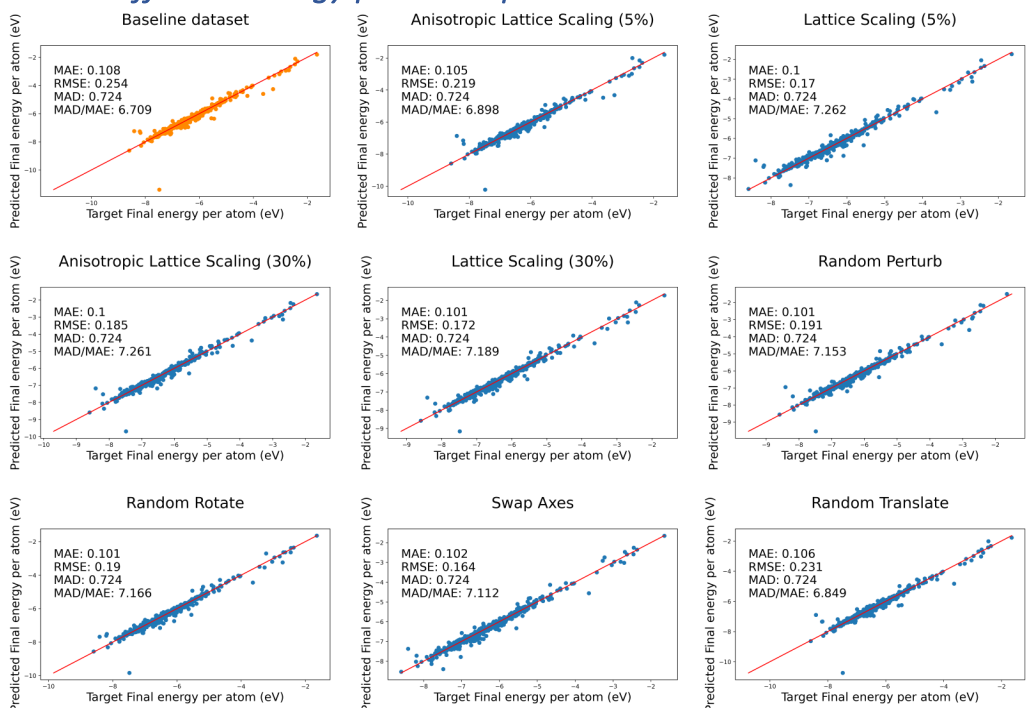

***Figure S9:** Comparison between different data augmentation strategies for energy per atom prediction. The relationship between the predicted and real values of the property is shown.*

## Figure S10: Different Fermi energy predictions.

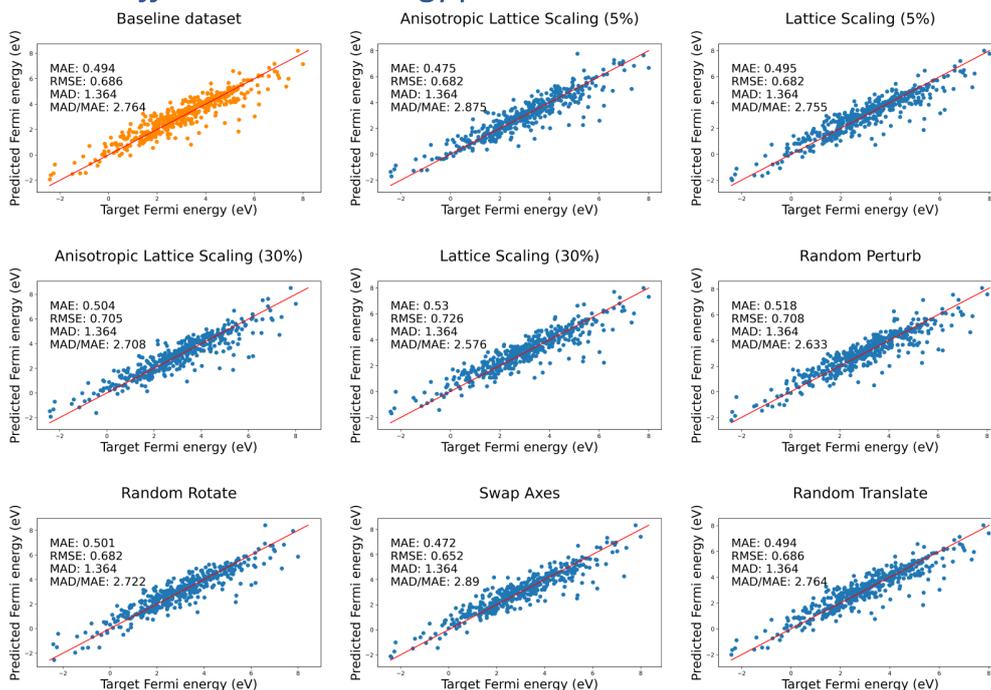

*Figure S10: Comparison between different data augmentation strategies for Fermi energy prediction. The relationship between the predicted and real values of the property is shown.*

## Figure S11: Different Band gap energy predictions.

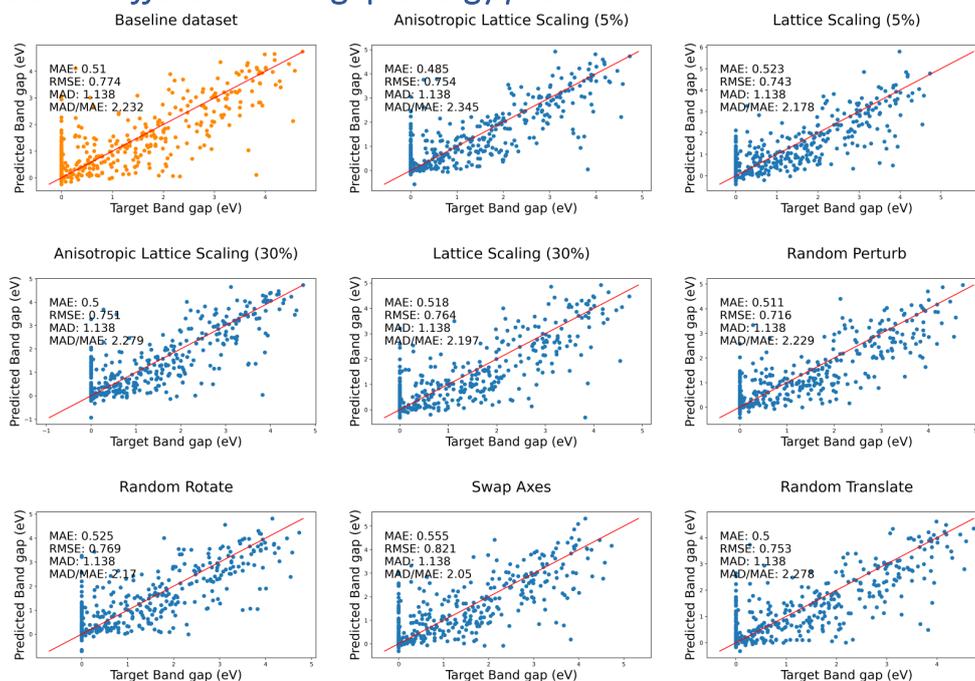

*Figure S11: Comparison between different data augmentation strategies for the band gap energy prediction. The relationship between the predicted and real values of the property is shown.*

*Figure S12: Isotropic lattice scaling: 5 %*

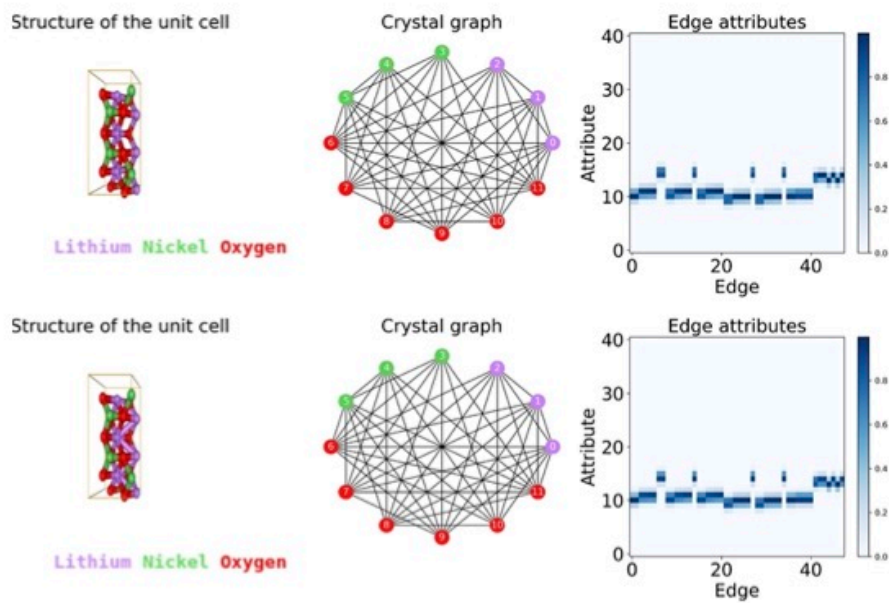

*Figure S12: Effect of isotropic lattice scaling (maximum volume change: 5%) on the initial material's structure and the corresponding CGCNN crystal graph. Top: original structure (LiNiO2, mp-866271); down: transformed structure (volume variation: 2.9%). Differences tend to appear in the edge attributes of crystal graph.*

*Figure S13: Isotropic lattice scaling: 30 %*

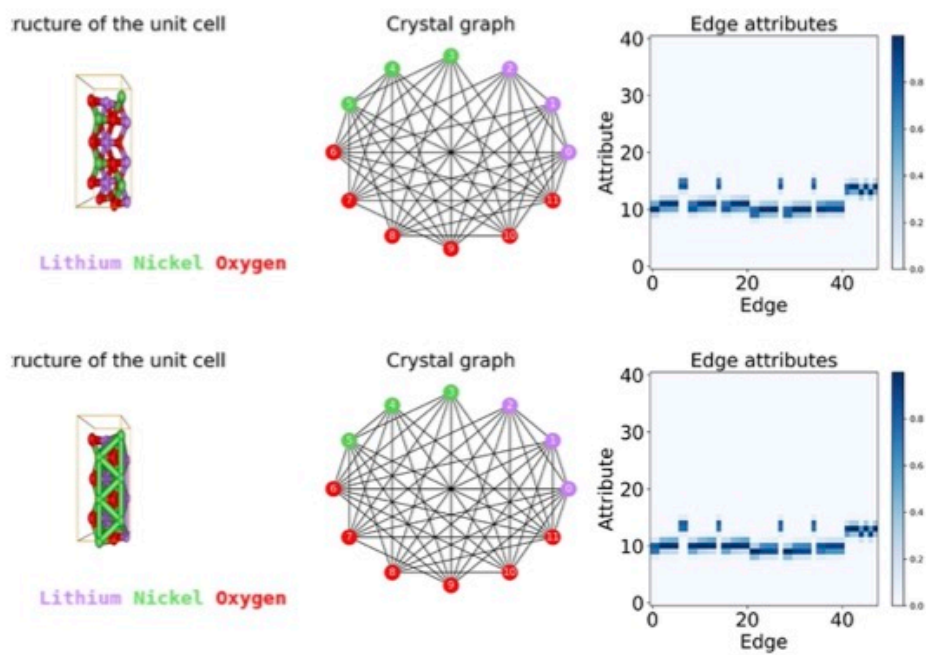

*Figure S13: Effect of isotropic lattice scaling (maximum volume change: 30%) on the initial material's structure and the corresponding CGCNN crystal graph. Top: original structure (LiNiO2, mp-866271); down: transformed structure (volume variation: 17.3%). Differences tend to appear in the edge attributes of crystal graph.*

*Figure S14: Anisotropic lattice scaling: 5%)*

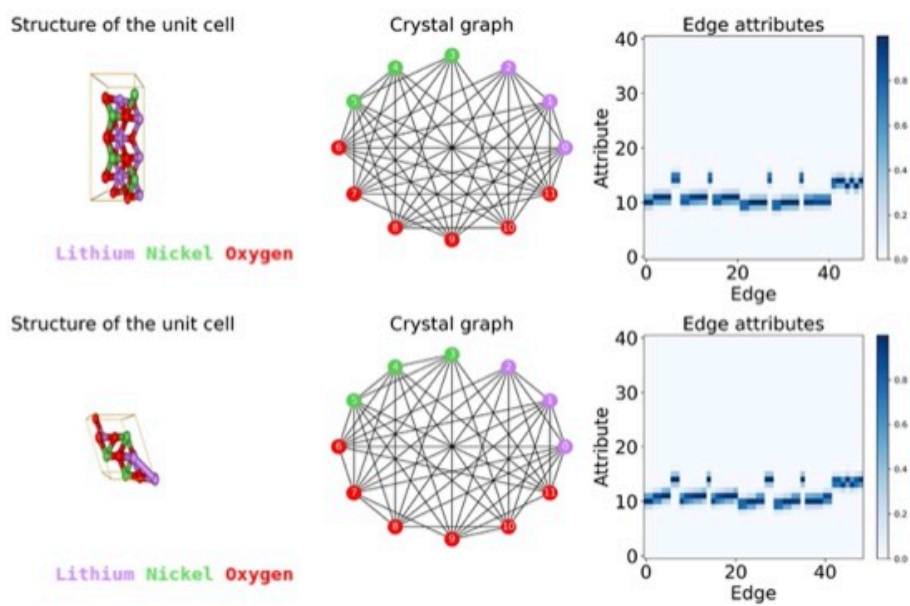

*Figure S14: Effect of anisotropic lattice scaling (maximum volume change: 5%) on the initial material's structure and the corresponding CGCNN crystal graph. Top: original structure (LiNiO2, mp-866271); down: transformed structure (volume variation: 4.5%). Differences tend to appear in the edge attributes of crystal graph.*

*Figure S15: Anisotropic lattice scaling: 30%*

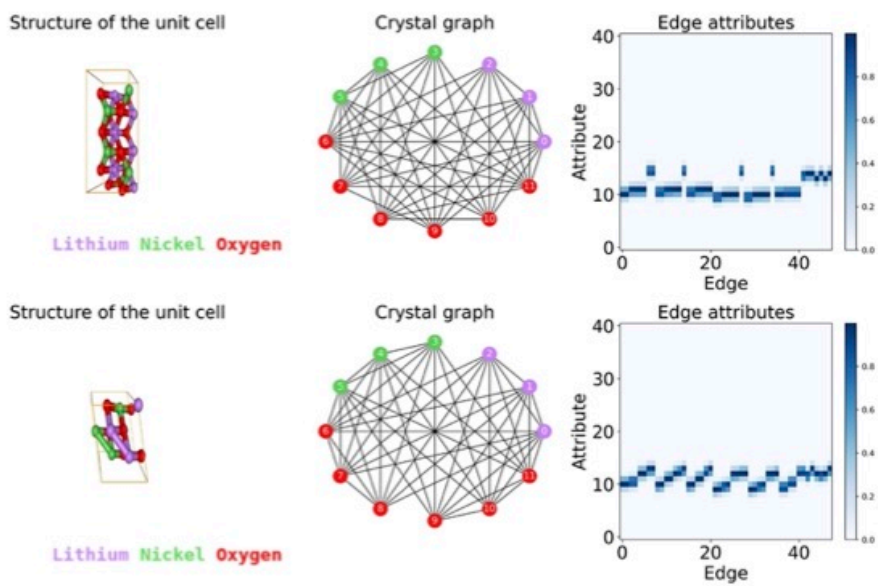

*Figure S15: Effect of anisotropic lattice scaling (maximum volume change: 30%) on the initial material's structure and the corresponding CGCNN crystal graph. Top: original structure (LiNiO$_2$, mp-866271); down: transformed structure (volume variation: 26%). Differences tend to appear in the edge attributes of crystal graph.*

## Figure S16: Comparison between the MAD/MAE ratios.

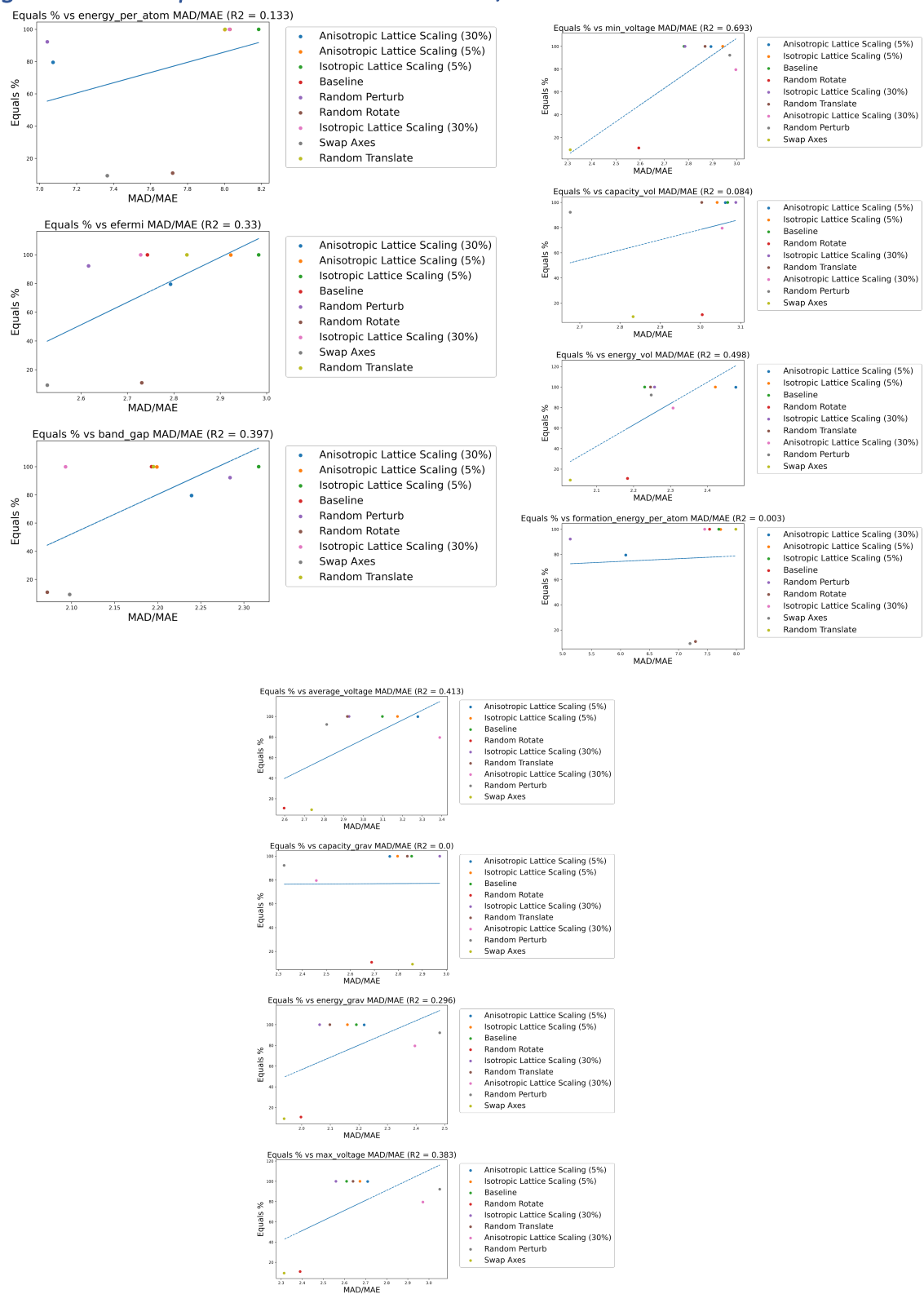

**Figure S16:** Comparison between the MAD/MAE ratio of each augmentation technique and the percentage of augmented structures that are considered equal to the original structures. Calculated with StructureMatcher method from pymatgen with all parameters set to default. Results from the first replicate.

*Figure S17: Comparison between the MAD/MAE ratios.*

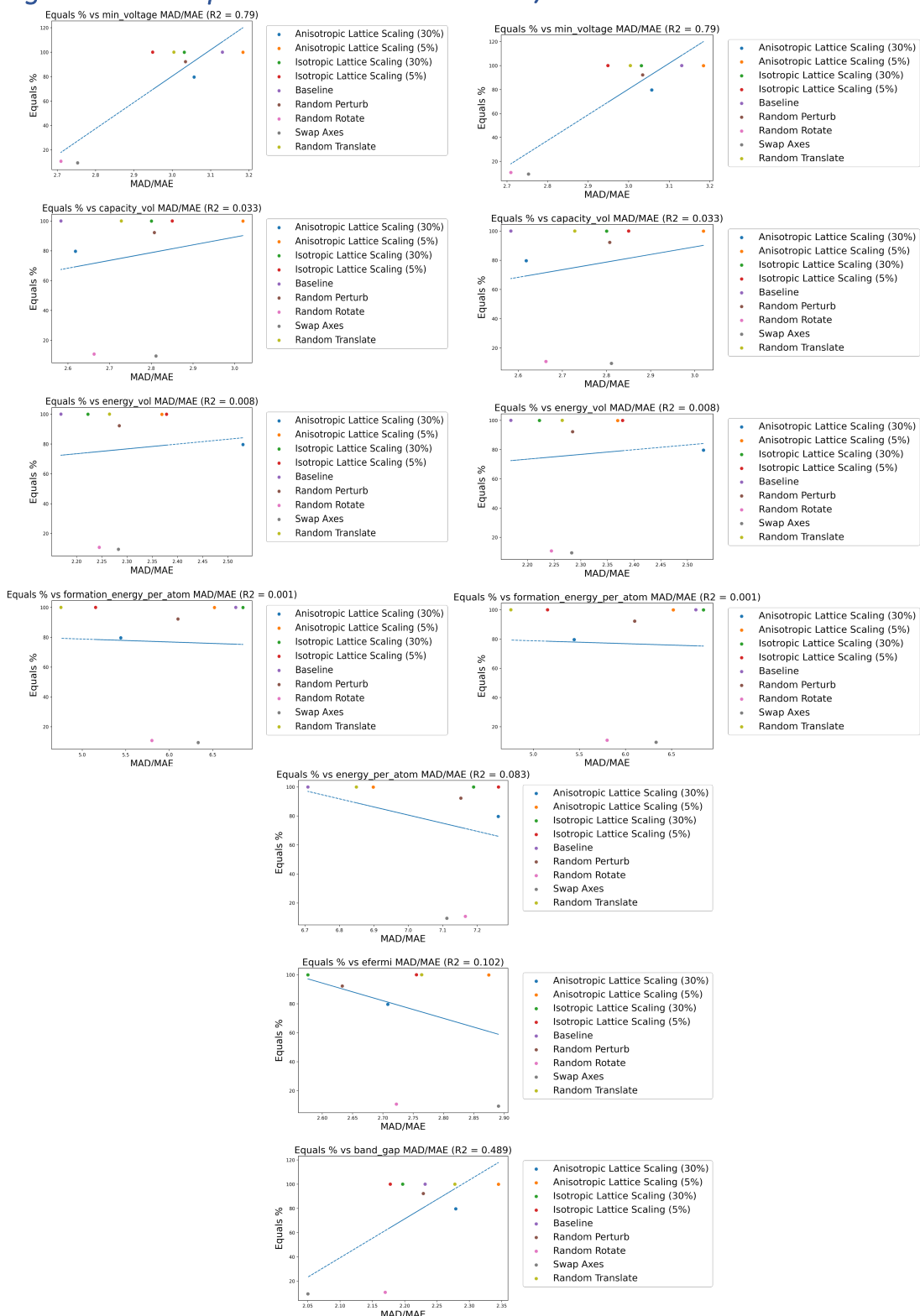

**Figure S17: Comparison between the MAD/MAE ratio of each augmentation technique and the percentage of augmented structures that are considered equal to the original structures. Calculated with StructureMatcher method from pymatgen with all parameters set to default. Results from the second replicate.**

## Figure S18: Comparison between the MAD/MAE ratios.

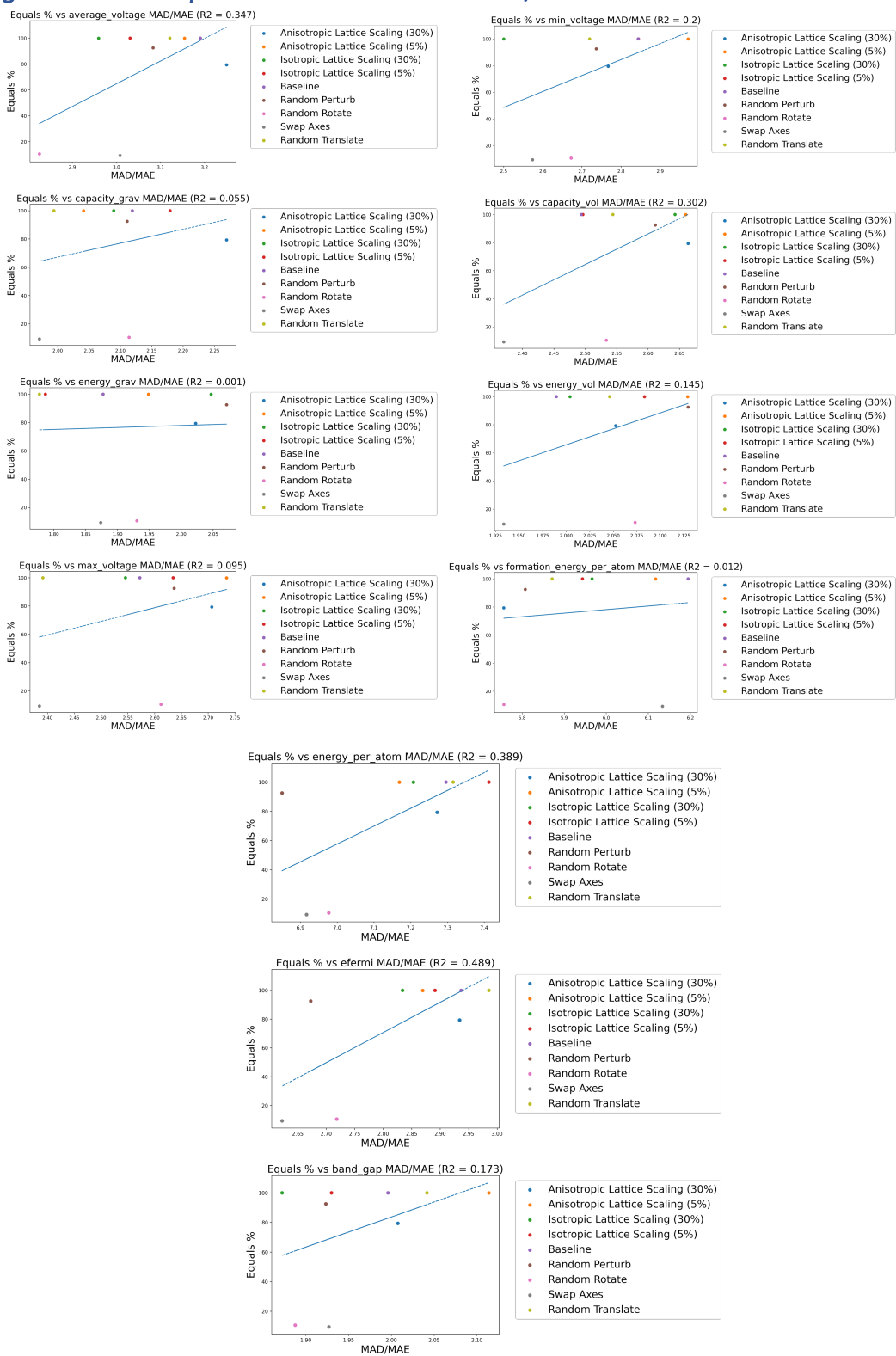

**Figure S18:** *Comparison between the MAD/MAE ratio of each augmentation technique and the percentage of augmented structures that are considered equal to the original structures. Calculated with StructureMatcher method from pymatgen with all parameters set to default. Results from the third replicate.*

*Figure S19: Comparison between the MAD/MAE ratios.*

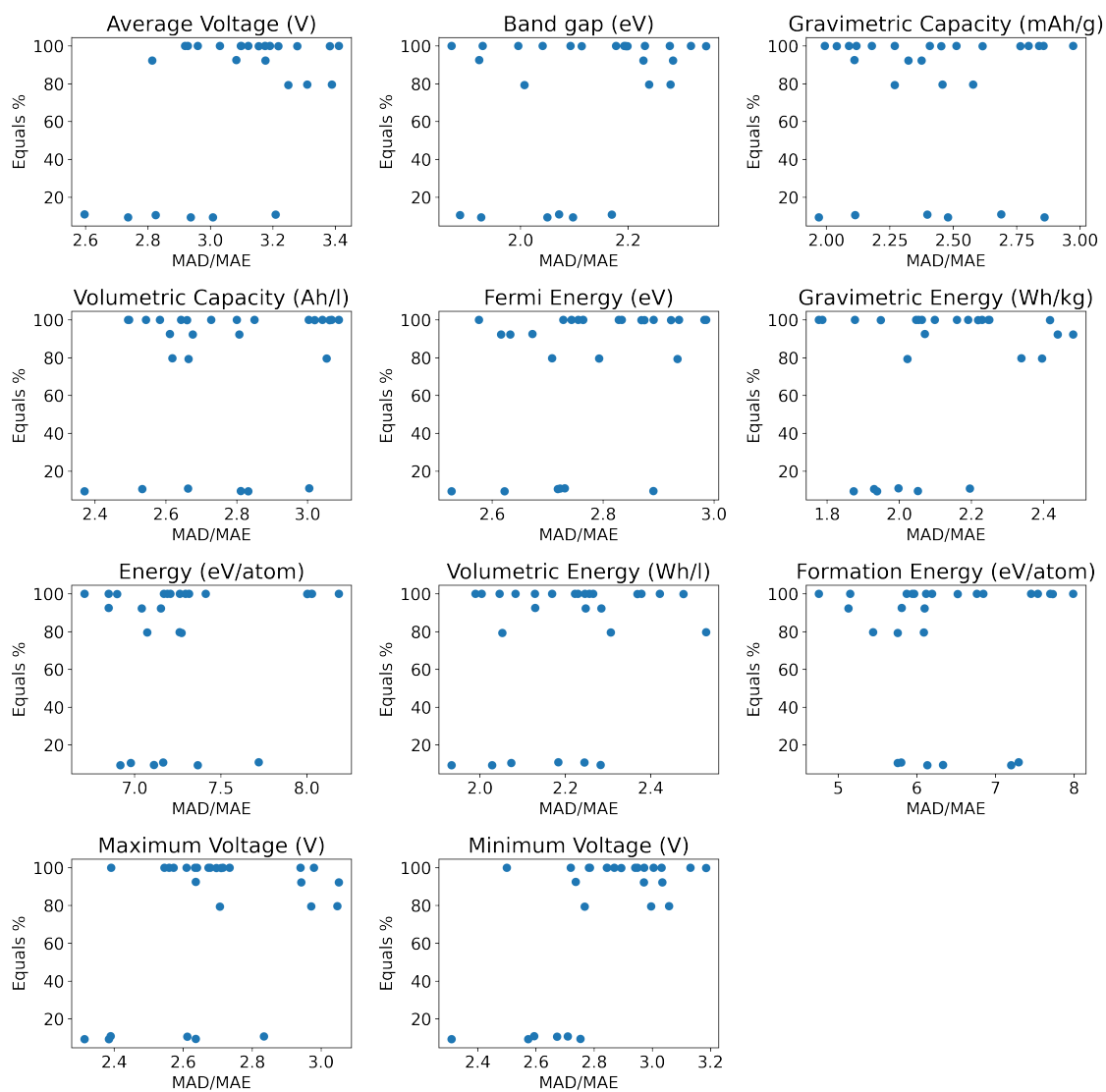

*Figure S19: Comparison between the MAD/MAE ratio of each augmentation technique and the percentage of augmented structures that are considered equal to the original structures. Calculated with StructureMatcher method from pymatgen with all parameters set to default. Results from all replicates.*